\documentclass[aps,prl,reprint,superscriptaddress,nofootinbib,longbibliography]{revtex4-2}

\usepackage{amsmath,amssymb,amsfonts,bm,mathtools}
\usepackage{graphicx}
\usepackage{xcolor}
\usepackage{hyperref}
\usepackage{ifthen}

\newcommand{\safeincludegraphics}[2][]{%
\IfFileExists{#2}{\includegraphics[#1]{#2}}{%
\fbox{\begin{minipage}{0.92\linewidth}
\centering
Missing figure file: \texttt{\detokenize{#2}}
\end{minipage}}%
}}

\hypersetup{
    colorlinks=true,
    linkcolor=blue,
    citecolor=blue,
    urlcolor=blue
}

\usepackage{hyperref} %
\usepackage{orcidlink}
\begin{document}

\title{Hidden Fisher Information in Many-Body Quantum Sensors}

\author{Atta ur Rahman~\!\!\orcidlink{0000-0001-7058-5671}}%\email{attapk@outlook.com}
\affiliation{School of Physical Sciences, University of Chinese Academy of Sciences, Yuquan Road 19A, Beijing 100049, China}

\author{Muhammad Noman~\!\!\orcidlink{0009-0003-6136-0114}}%\email{mnoman90k@gmail.com}
\affiliation{Institute of Fundamental Physics and Quantum Technology and School of Physical Science and Technology, Ningbo University, Ningbo 315211, China}

\author{Samira Elghaayda~\!\!\orcidlink{0000-0002-6655-0465}}
%\email{samira.elghaayda@kfupm.edu.sa}
\affiliation{Interdisciplinary Research Center for Advanced Quantum Computing, King Fahd University of Petroleum and Minerals, Dhahran 31261, Saudi Arabia}

\author{Saeed Haddadi~\!\!\orcidlink{0000-0002-1596-0763}}\email{haddadi@ipm.ir}
\affiliation{School of Particles and Accelerators, Institute for Research in Fundamental Sciences (IPM), P.O. Box 19395-5531, Tehran, Iran}

\date{\today}

\begin{abstract}
Quantum Fisher information (QFI) determines the ultimate precision of a quantum sensor after optimization over all allowed measurements, but it does not quantify how difficult the optimal readout is to implement. We introduce a moment Fisher hierarchy \(F_M^{(k)}\), which measures the fraction of QFI accessible to collective spin moments up to degree \(k\), together with a readout-complexity scale \(k_{90}\) defined by the minimum degree required to recover \(90\%\) of the QFI. Applying this framework to a nonlinear collective-spin sensor, we show that many-body dynamics can preserve large QFI while transferring the optimal readout into high-rank \(SU(2)\) tensor sectors inaccessible to low-order collective observables. For system sizes up to \(N=100\), first moments access only \(\sim 10^{-2}\) of the available QFI, while the degree required to recover \(90\%\) of it grows to \(k_{90}\simeq 28\). A rank-resolved analysis of the symmetric logarithmic derivative (SLD) under weak damping shows that the nonlinear protocol pushes the optimal readout toward the highest available tensor ranks while strongly depleting its low-rank weight. These results identify nonlinear tangent-space shear as the mechanism behind a finite-size readout-complexity bottleneck in many-body quantum metrology.
\end{abstract}

\maketitle

\textit{Introduction.---}
Quantum-enhanced metrology aims to measure weak signals with a precision that cannot be reached by independent particles alone \cite{Giovannetti2004, Giovannetti2011}. The central idea is to use nonclassical many-body states---for example, spin-squeezed \cite{Wineland1992, Kitagawa1993, Ma2011}, strongly entangled \cite{Toth2014, Pezze2018, Braun2018}, critical \cite{Zanardi2006}, or dynamically correlated states \cite{Frowis2015, Fiderer2018}---as probes whose response to an unknown parameter is amplified by quantum correlations. In this setting, the quantum Fisher information (QFI) has become the standard theoretical benchmark \cite{Paris2009, Degen2017}. Given a parameter $\alpha$ encoded in a probe state $\rho_\alpha$, the QFI $F_Q$ determines the best possible local sensitivity through the quantum Cram\'er--Rao bound \cite{Helstrom1976, Holevo1982, Braunstein1994}. It tells us how much metrological information is contained in the quantum state when the measurement is allowed to be fully optimized \cite{Giovannetti2006}.

Although the QFI gives the ultimate achievable precision, it does not quantify the complexity of the measurement required to attain that precision \cite{Dooley2018, Meyer2021}. In many-body systems, the symmetric logarithmic derivative (SLD) defining the optimal readout can become a highly nonlocal operator \cite{Zhuang2018}. It often retains little overlap with experimentally accessible observables such as collective magnetization or low-order correlations \cite{Pezze2014, Gessner2019b}. A many-body sensor may therefore possess large QFI while remaining operationally difficult to read out.

This readout issue has become increasingly relevant because experimental platforms have made remarkable progress in preparing correlated many-body probes. Metrologically useful correlations have been generated across multiple highly controlled physical systems. For example, in cavity-QED ensembles, light-matter interactions create robust entangled states \cite{Leroux2010, Hosten2016, Bao2020}, with recent advances demonstrating complex programmable interactions \cite{Pedrozo2020, Greve2022, Robinson2024}. Similar metrological potential has been unlocked in trapped ions \cite{Bohnet2016, Marciniak2022, LewisSwan2024} and spinor condensates \cite{Linnemann2016, Mao2023}. Concurrently, Rydberg arrays and optical clocks have shown that strong dipole interactions can generate states with immense metrological capability \cite{Bornet2023, Eckner2023, Cao2024, Bluvstein2024, Schine2024}. These systems show that many-body interactions can create states with strong metrological potential. At the same time, they also show that state preparation alone is not sufficient: the experimentally observed sensitivity often depends crucially on the final readout.

This realization has motivated several advanced measurement strategies. Interaction-based readout and phase magnification transform complicated correlations into simpler, readable signals \cite{Strobel2014, Davis2016, Hosten2016b, Nolan2017}. Time reversal effectively disentangles the probe prior to detection \cite{Macri2016, Colombo2022}. Meanwhile, parity-sensitive detection and number-resolved measurements extract fine-grained quantum features \cite{Gessner2019, Kaubruegger2021}. Furthermore, randomized measurement protocols and variational approaches reconstruct state properties without demanding full tomography \cite{Cerezo2021, Elben2023, Choi2023, Oh2024, Haug2024, Vitale2024}. These methods can be understood as ways of decoding, amplifying, or reconstructing many-body information before it is finally measured. Their success points to a basic question: when is such an advanced readout merely helpful, and when is it necessary? This question is not only technical, but also interpretational.  In many experiments, the same observed sensitivity can be limited either by imperfect state preparation or by an insufficiently expressive measurement.  Without a diagnostic that separates these two possibilities, a state with large hidden metrological response may be mistaken for a state with little metrological value.  A central aim of the present work is to make this distinction operational.

At present, this question lacks a compact degree-resolved diagnostic. The QFI gives the best possible sensitivity over all measurements. The classical Fisher information gives the sensitivity of one specified measurement. Between these two extremes, we need a tool that quantifies how much of the QFI is accessible when the experiment is restricted to a controlled family of simple observables. For collective spin sensors, a natural family is given by collective moments. First moments describe ordinary collective magnetization. Second moments include spin-squeezing and covariance measurements \cite{Gessner2019b}. Higher moments probe increasingly complex many-body correlations. A hierarchy organized by moment degree therefore provides a direct measure of readout complexity.

Existing approaches typically characterize either the ultimate metrological sensitivity through the QFI or the performance of a particular measurement protocol. By contrast, the present hierarchy resolves how metrological information is distributed across collective operator complexity. This identifies a distinct notion of hidden Fisher information: information that remains present in the quantum state but is inaccessible to low-degree collective readout. The hierarchy therefore provides a quantitative bridge between metrological sensitivity and experimentally accessible measurement complexity.

Here we introduce such a hierarchy. For each integer $k$, we define the moment Fisher information $F_M^{(k)}$ as the maximum error-propagation Fisher information obtainable from scalar estimators built from collective spin polynomials of degree at most $k$. The normalized quantity
\begin{equation}
\eta_k=\frac{F_M^{(k)}}{F_Q}
\label{eq:etak_intro}
\end{equation}
measures the fraction of the QFI accessible to collective moments up to degree $k$. We also define the readout-complexity scale
\begin{equation}
k_{90}=\min\{k:\,F_M^{(k)}\geq 0.9F_Q\},
\label{eq:k90_intro}
\end{equation}
which is the minimum collective moment degree required to recover $90\%$ of the QFI. A small $k_{90}$ means that the metrological information is directly visible to low-order collective readout. A large, system-size-dependent $k_{90}$ means that the state contains useful information, but that the information is distributed over high-order collective correlations.

The moment Fisher hierarchy complements, rather than replaces, the QFI. The QFI asks what is possible if the measurement is unrestricted. The hierarchy asks what remains accessible when the readout is restricted to collective moments of bounded degree. This distinction is important because many experiments do not directly implement arbitrary many-body measurements. They typically access low-order collective observables, possibly after global rotations, unless additional resources such as time reversal, interaction-based decoding, number resolution, randomized measurements, or tomography are supplied. Thus, $F_M^{(k)}$ provides a practical way to diagnose whether a sensor is limited by state preparation or by readout complexity.

To demonstrate the hierarchy in a controlled many-body setting, we study a kicked collective spin, corresponding to a symmetric-subspace realization of the quantum kicked top \cite{Haake1987, Chaudhury2009}. The kicked top provides a minimal nonlinear many-body sensor \cite{Zarum1998, Wang2004, Sieberer2019} in which the full collective operator hierarchy can be resolved at finite $N$. It provides a tunable non-Gaussian framework where the QFI can be compared directly with accessible moment Fisher information, and the structure of the optimal readout can be resolved in operator space.

We find a sharp separation between information storage and information accessibility. In the linear protocol, the QFI is almost completely accessible through first moments. In the resonant nonlinear protocol, by contrast, the QFI remains present but becomes largely invisible to low-order collective observables. For system sizes up to $N=100$, first moments access only about $10^{-2}$ of the QFI, while the degree required to recover $90\%$ of the information grows to $k_{90}\simeq 28$. The endpoint of the hierarchy recovers the QFI, confirming that the information is not lost. Instead, it is redistributed into high-degree collective moment sectors.

We identify the physical mechanism as nonlinear tangent-space shear. The parameter derivative of the state initially corresponds to a smooth displacement on the collective Bloch sphere. Strong nonlinear kicking stretches this tangent direction into fine angular structures. In operator language, the associated optimal readout, represented locally by the SLD, acquires substantial weight in high-rank irreducible $SU(2)$ tensor sectors \cite{Dowling1994, McConnell2015}. Since collective moments of degree $k$ resolve tensor ranks only up to $\ell=k$, this redistribution makes the optimal readout progressively inaccessible to low-degree moments. This gives a direct geometric interpretation of hidden Fisher information: the state is sensitive to the parameter, but the observable that reveals this sensitivity has become a high-rank many-body operator. Because the mechanism relies on nonlinear generation of high-rank collective operator structure, it is expected to arise broadly in non-Gaussian many-body sensors beyond the kicked-top setting.

Our contribution is therefore threefold. First, we introduce a finite-$N$ moment Fisher hierarchy that quantifies how much QFI is accessible to collective moment readout of a given degree. Second, we define $k_{90}$ as an operational readout-complexity scale for many-body sensors. Third, we show that nonlinear many-body dynamics can preserve QFI while moving the optimal readout into high-rank operator sectors, thereby making the information hidden from standard low-order measurements. This establishes hidden Fisher information as a concrete measurement-complexity bottleneck in many-body quantum metrology.

\textit{Model.---}
We consider $N$ spin-$1/2$ particles restricted to the permutation-symmetric manifold with total spin $j=N/2$. One Floquet step is defined by
\begin{equation}
U_{\alpha,\beta}
=
e^{-i\alpha J_z}
e^{-i\beta J_y^2/(2j)} ,
\label{eq:Floquet}
\end{equation}
where $\alpha$ is the unknown rotation angle to be estimated and $\beta$ controls the strength of a collective one-axis-twisting kick. The initial probe is the spin coherent state
\begin{equation}
|\Theta,\Phi\rangle
=
e^{i\Theta(J_x\sin\Phi-J_y\cos\Phi)}
|j,j\rangle ,
\qquad
\Theta=\pi/3,\quad \Phi=0 .
\label{eq:coherent}
\end{equation}
Unless otherwise stated, the parameter is evaluated locally at $\alpha=\pi/2$, and the large-$N$ moment-hierarchy calculation uses eight unitary kicks. We compare a linear baseline, $\beta=0$, with a resonant nonlinear protocol, $\beta=\pi j$.

\begin{figure*}[t]
\centering
\safeincludegraphics[width=0.98\textwidth]{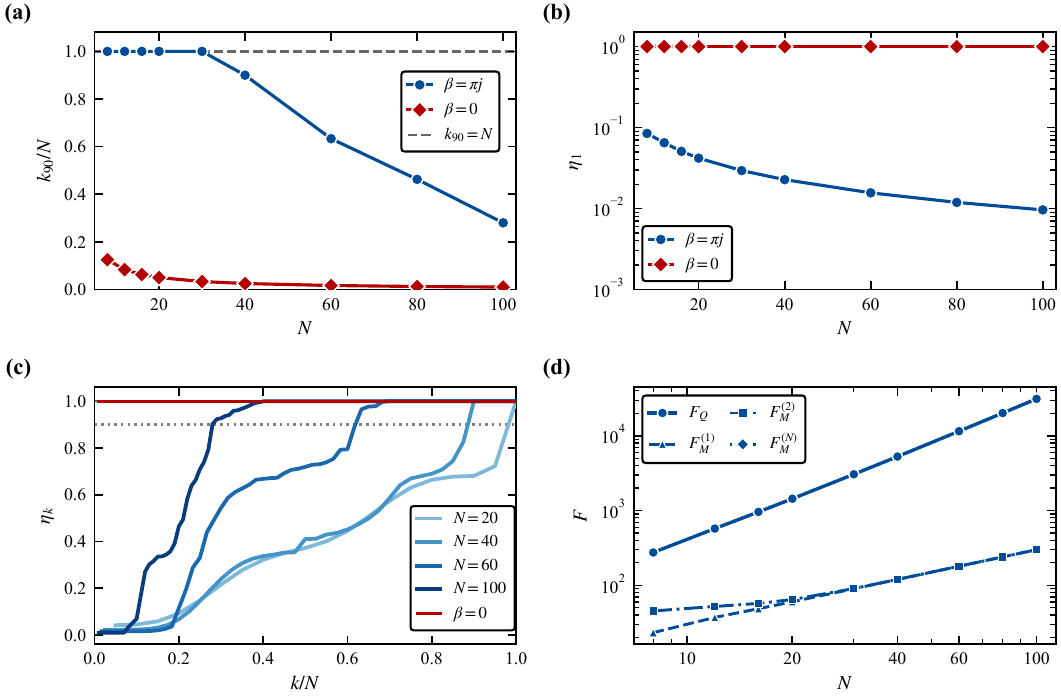}
\caption{
\textbf{Large-$N$ moment-readout hierarchy for the kicked-top sensor.}
(a) Normalized readout depth $k_{90}/N$. The linear protocol, $\beta=0$, requires only the lowest moment degree, whereas the resonant nonlinear protocol, $\beta=\pi j$, retains a much larger readout depth over the simulated range.
(b) First-moment accessibility $\eta_1=F_{\rm M}^{(1)}/F_Q$. For the nonlinear protocol, $\eta_1$ falls to the percent level by $N=100$, while the linear protocol remains near unity.
(c) Cumulative accessibility $\eta_k=F_{\rm M}^{(k)}/F_Q$ versus scaled moment degree $k/N$ for $N=20,40,60,$ and $100$ in the nonlinear protocol; the $\beta=0$ curve at $N=100$ is shown as a low-complexity reference. The dotted line marks $\eta_k=0.9$.
(d) Absolute Fisher-information hierarchy for the nonlinear protocol: $F_Q$, $F_{\rm M}^{(1)}$, $F_{\rm M}^{(2)}$, and $F_{\rm M}^{(N)}$. The endpoint $F_{\rm M}^{(N)}\simeq F_Q$ confirms completeness, whereas the large gap to $F_{\rm M}^{(1)}$ and $F_{\rm M}^{(2)}$ shows that the available QFI is hidden from low-degree moment readout.
}
\label{fig:moment_hierarchy}
\end{figure*}

\begin{figure*}[t]
\centering
\safeincludegraphics[width=0.92\textwidth]{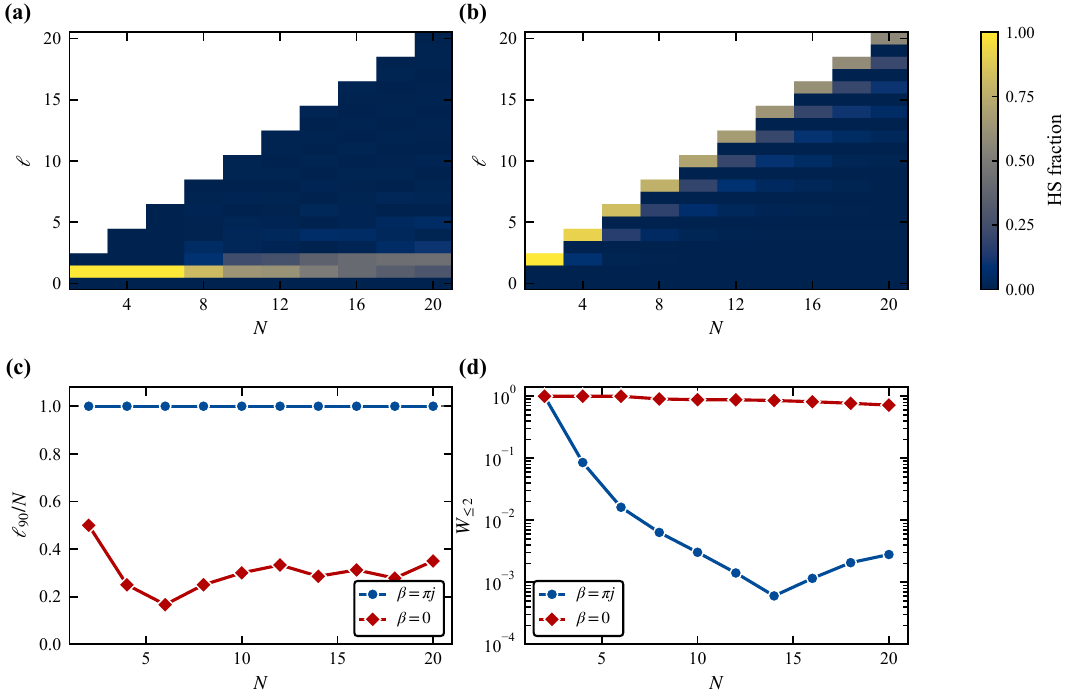}
\caption{
\textbf{The optimal SLD readout is transferred to high tensor rank.}
Rank-resolved SLD diagnostics for one weakly damped Floquet step with $\gamma=10^{-3}$ and $N\leq20$.
(a) Hilbert--Schmidt rank fraction $w_\ell(L_\alpha)$ for the linear protocol, $\beta=0$.
(b) The same quantity for the resonant nonlinear protocol, $\beta=\pi j$. The weight shifts toward the upper edge of the allowed tensor-rank range.
(c) Normalized 90\%-weight rank $\ell_{90}/N$. In the nonlinear protocol, capturing $90\%$ of the SLD weight requires the maximal available rank throughout the displayed range, whereas the linear protocol remains concentrated at substantially lower rank.
(d) Low-rank SLD weight $W_{\leq2}$. The nonlinear protocol suppresses the low-rank contribution to the $10^{-3}$ scale, while the linear protocol retains order-unity weight in these sectors. The color bar in (a,b) denotes the Hilbert--Schmidt fraction.
}
\label{fig:rank_structure}
\end{figure*}

The large-$N$ hierarchy in Fig.~\ref{fig:moment_hierarchy} is evaluated under unitary dynamics, which allows the calculation to reach $N=100$. For the rank-resolved SLD diagnostic in Fig.~\ref{fig:rank_structure}, we additionally include weak collective damping and evaluate one noisy Floquet step for $N\leq20$,
\begin{equation}
{\cal L}_d(\rho)
=
\gamma\left(2J_-\rho J_+ - \{J_+J_-,\rho\}\right),
\qquad
\gamma=10^{-3},
\label{eq:damping}
\end{equation}
implemented as
\begin{equation}
\rho \mapsto
U_{\alpha,\beta}
\left[
e^{{\cal L}_d}\rho
\right]
U_{\alpha,\beta}^{\dagger}.
\label{eq:noisy_map}
\end{equation}
This mixed-state calculation provides an independent check that the high-rank readout structure is not restricted to ideal pure-state evolution.

For a final state
\begin{equation}
\rho_\alpha=\sum_a p_a |a\rangle\langle a|,
\end{equation}
the QFI is
\begin{equation}
F_Q
=
2\sum_{a,b}
\frac{
|\langle a|\partial_\alpha\rho_\alpha|b\rangle|^2
}{
p_a+p_b
},
\label{eq:QFI}
\end{equation}
where terms with $p_a+p_b=0$ are omitted \cite{Gorecki2020, Liu2014, Alipour2014}. Equivalently,
\begin{equation}
F_Q=\mathrm{Tr}(\rho_\alpha L_\alpha^2),
\end{equation}
where the SLD $L_\alpha$ satisfies
\begin{equation}
\partial_\alpha\rho_\alpha
=
\frac{1}{2}
\left(
L_\alpha\rho_\alpha+\rho_\alpha L_\alpha
\right).
\label{eq:SLD}
\end{equation}
The SLD specifies the locally optimal measurement direction. Our central question is whether this direction has substantial projection onto experimentally simple collective moments.

\textit{Moment Fisher hierarchy.---}
Let ${\cal M}_k$ be the real vector space spanned by traceless Hermitian symmetrized products of collective spin components up to degree $k$:
\begin{equation}
{\cal M}_k
=
\mathrm{span}_{\mathbb{R}}
\left\{
(J_{a_1}\cdots J_{a_q})_{\mathrm{sym}}
:
1\leq q\leq k,\;
a_i\in\{x,y,z\}
\right\},
\label{eq:Mk}
\end{equation}
after removing identity components and linear dependencies. For a Hilbert--Schmidt orthonormal basis $\{O_\mu\}$ of ${\cal M}_k$, define the response vector
\begin{equation}
g_\mu
=
\mathrm{Tr}
\left[
(\partial_\alpha\rho_\alpha)O_\mu
\right],
\label{eq:g}
\end{equation}
and the covariance matrix
\begin{equation}
\Sigma_{\mu\nu}
=
\frac{1}{2}
\mathrm{Tr}
\left[
\rho_\alpha
\{\Delta O_\mu,\Delta O_\nu\}
\right],
\qquad
\Delta O_\mu=O_\mu-\langle O_\mu\rangle .
\label{eq:Sigma}
\end{equation}
For a scalar moment estimator
\begin{equation}
X=\sum_\mu c_\mu O_\mu,
\end{equation}
the error-propagation Fisher information is
\begin{equation}
F_X
=
\frac{
(\partial_\alpha\langle X\rangle)^2
}{
(\Delta X)^2
}
=
\frac{
(c^Tg)^2
}{
c^T\Sigma c
}.
\label{eq:FX}
\end{equation}
Optimizing over all real coefficients $c$ gives
\begin{equation}
F_M^{(k)}
=
g_k^T\Sigma_k^+g_k,
\qquad
\eta_k=
\frac{F_M^{(k)}}{F_Q},
\label{eq:FMk}
\end{equation}
where $\Sigma_k^+$ is the Moore--Penrose pseudoinverse.

The quantity $F_M^{(k)}$ should be interpreted carefully. It is not the maximum Fisher information over all POVMs compatible with a given apparatus. Rather, it is the maximum information accessible to scalar estimators constructed from collective moments up to degree $k$. This distinction is important. Number-resolved histograms, parity-sensitive measurements, tomography, randomized measurements, or interaction-based decoding can access information beyond $F_M^{(k)}$ at fixed low $k$ because they implicitly use richer measurement resources. The purpose of the hierarchy is therefore not to prove that all simple collective measurements fail, but to quantify precisely what is visible to low-degree moment readout.

At finite $N$, the hierarchy is complete. In the spin-$j=N/2$ irreducible representation, collective spin polynomials up to degree $N$ span the full traceless Hermitian operator space,
\begin{align}
{\cal M}_N
=
\mathrm{Herm}_0({\cal H}_j)
\equiv
\left\{
O\in{\cal B}({\cal H}_j):
O=O^\dagger,\ \mathrm{Tr}\,O=0
\right\},
\label{eq:MN_complete}
\end{align}
Here $j=N/2 $, and ${\cal H}_j$ is the spin-$j$ symmetric Hilbert space and $\mathrm{Herm}_0({\cal H}_j)$ denotes the real vector space of traceless Hermitian operators acting on it. Since the SLD is contained in this space, the endpoint of the hierarchy satisfies
\begin{equation}
F_M^{(N)}=F_Q
\label{eq:endpoint}
\end{equation}
within numerical precision. This endpoint recovery is a crucial diagnostic: it distinguishes hidden Fisher information from numerical loss of QFI.

\textit{Tangent-space shear mechanism.---}
The mechanism responsible for hiding the information can be understood by propagating the parameter tangent
\begin{equation}
X_t=\partial_\alpha\rho_t.
\end{equation}
Differentiating the noisy Floquet map gives
\begin{equation}
X_{t+1}
=
{\cal U}_{\alpha,\beta}
[
{\cal D}(X_t)
]
-i[J_z,\rho_{t+1}],
\label{eq:tangent}
\end{equation}
where ${\cal D}=e^{{\cal L}_d}$ and
\begin{equation}
{\cal U}_{\alpha,\beta}(A)
=
U_{\alpha,\beta}AU_{\alpha,\beta}^{\dagger}.
\end{equation}
The source term $-i[J_z,\rho_{t+1}]$ is generated by the parameter rotation. The complexity enters through the repeated adjoint action of the nonlinear kick.

For $\beta=0$, the dynamics is a rigid collective rotation. The tangent direction remains smooth on the collective Bloch sphere and is well captured by low-rank collective observables. By contrast, for the nonlinear kick, the corresponding classical map on the normalized spin vector $s=J/j$ is
\begin{align}
s_y^+ &= s_y,\\
s_x^+ &= s_x\cos(\beta s_y)-s_z\sin(\beta s_y),\\
s_z^+ &= s_z\cos(\beta s_y)+s_x\sin(\beta s_y).
\end{align}
The local twist angle is
\begin{equation}
\chi(s)=\beta s_y.
\end{equation}
At resonance, $\beta=\pi j$, and therefore
\begin{equation}
\frac{\partial\chi}{\partial s_y}
=
\beta
=
O(N).
\label{eq:shear}
\end{equation}
A smooth parameter displacement is then strongly sheared and stretched into angular structures whose characteristic scale decreases as $1/N$. Quantum mechanically, this corresponds to the migration of the tangent vector and the SLD into high-rank irreducible tensor sectors \cite{Dowling1994}:
\begin{equation}
e^{i\beta J_y^2/(2j)}
T_{\ell m}
e^{-i\beta J_y^2/(2j)}
=
\sum_{\ell' m'}
C^{\ell m}_{\ell' m'}(\beta,j)
T_{\ell' m'} .
\label{eq:tensor_mixing}
\end{equation}
Since degree-$k$ collective moments resolve tensor ranks only up to $\ell=k$, this shear mechanism predicts a growing finite-size readout depth as the parameter response is transferred into progressively higher tensor-rank sectors.

\textit{Hidden Fisher information.---}
Figure~\ref{fig:moment_hierarchy} presents the central large-$N$ result. For the linear baseline $\beta=0$, the QFI is operationally simple: first moments nearly saturate the QFI and $k_{90}=1$. Consequently, $k_{90}/N$ decreases as the system grows, as expected for a readout whose absolute degree remains fixed.

The resonant nonlinear protocol behaves very differently. Although the QFI grows strongly with $N$, first-moment accessibility decreases to the percent level, reaching $\eta_1\simeq10^{-2}$ at $N=100$. Panel (d) shows simultaneously that $F_{\rm M}^{(N)}$ tracks $F_Q$, whereas $F_{\rm M}^{(1)}$ and $F_{\rm M}^{(2)}$ remain far below it. The gap therefore reflects inaccessible readout structure rather than loss of QFI.

The readout degree required to recover most of the QFI also grows substantially in absolute terms. At $N=100$, the data give $k_{90}\simeq28$, corresponding to $k_{90}/N\simeq0.28$. The hierarchy curves in panel (c) show that the recovery is distributed over many moment degrees rather than being exhausted by $k=1$ or $k=2$. These finite-size data establish a pronounced growth of readout depth over the simulated range; they are not used here to assume a specific asymptotic scaling law.

Figure~\ref{fig:rank_structure} provides an independent weak-damping mixed-state diagnostic up to $N=20$. The Supplemental Material further tests the robustness of the hierarchy against detuning and numerical choices. Together, these checks show that the observed separation between QFI and low-order accessibility is not an artifact of a single idealized numerical construction.

There is also an important interplay with decoherence. Dissipation acts as a low-pass filter on the collective Bloch sphere and suppresses high-rank tensor components more rapidly than low-rank components \cite{Escher2011, DemkowiczDobrzanski2012}. Therefore, in a strict asymptotic limit at fixed nonzero dissipation, ultra-high-rank information may eventually be degraded. The finite-size message is nevertheless operationally relevant: before decoherence erases the metrological advantage, a sensor can already enter a regime where substantial QFI exists but low-order collective moments access only a small fraction of it.

\textit{Optimal readout structure.---}
To verify that the barrier is intrinsic and not an artifact of a particular polynomial basis, we decompose the relevant operators into irreducible $SU(2)$ tensor ranks \cite{McConnell2015}. Let $T_{\ell m}$ be Hilbert--Schmidt orthonormal eigenoperators of the adjoint Casimir,
\begin{equation}
{\cal C}_{\mathrm{ad}}(O)
=
\sum_{a=x,y,z}
[J_a,[J_a,O]],
\qquad
{\cal C}_{\mathrm{ad}}(T_{\ell m})
=
\ell(\ell+1)T_{\ell m}.
\label{eq:Casimir}
\end{equation}
For an operator $O$, define the rank weight
\begin{equation}
w_\ell(O)
=
\frac{
\sum_{m=-\ell}^{\ell}
|\mathrm{Tr}(T_{\ell m}^{\dagger}O)|^2
}{
\mathrm{Tr}(O^{\dagger}O)
}.
\label{eq:rank_weight}
\end{equation}
For the SLD, we further define the cumulative 90\%-weight rank and the low-rank weight
\begin{equation}
\ell_{90}
=
\min\left\{
\ell:
\sum_{r=0}^{\ell} w_r(L_\alpha)\geq0.9
\right\},
\quad
W_{\leq2}
=
\sum_{\ell=0}^{2} w_\ell(L_\alpha).
\label{eq:rank_diagnostics}
\end{equation}
The quantity $\ell_{90}$ measures how far in tensor rank one must go to capture most of the SLD, while $W_{\leq2}$ measures how much of the locally optimal readout remains in the lowest-rank sectors.

Figure~\ref{fig:rank_structure} resolves the SLD directly in irreducible tensor rank. In the linear baseline, the SLD weight is concentrated near the lowest ranks, consistent with the success of low-degree moment readout. Under the resonant nonlinear kick, the heatmap shifts toward the upper boundary $\ell\sim N$. This difference is quantified in panel (c): over $N=2$--$20$, the nonlinear protocol has $\ell_{90}/N=1$, while the linear protocol reaches the same cumulative weight at much smaller normalized rank.

Panel (d) provides the complementary low-rank diagnostic. For $\beta=\pi j$, $W_{\leq2}$ is rapidly suppressed from order unity at the smallest system to the $10^{-3}$ scale over most of the studied range. By contrast, the linear baseline retains order-unity low-rank weight. Because the SLD specifies the locally optimal measurement direction, these two diagnostics give a basis-independent explanation of the hierarchy gap in Fig.~\ref{fig:moment_hierarchy}: the QFI remains available, but the readout that exposes it has been transferred out of the low-rank collective sector.

This observation clarifies the meaning of hidden Fisher information. The problem is not that the state lacks sensitivity to $\alpha$. The problem is that the direction in operator space that converts this sensitivity into a signal has little support in the experimentally simple low-degree sector. The QFI is present, but the readout needed to attain it is structurally complex.

\textit{Relation to richer collective measurements.---}
The moment hierarchy deliberately focuses on low-degree scalar moment estimators. It therefore does not exclude the possibility that other collective measurements recover more information. Indeed, the Supplemental Material compares the hierarchy with explicit collective readouts based on optimized global rotations followed by number-resolved $J_z$ histograms or parity measurements. These readouts can recover substantially more information than linear moments because a full histogram contains high-order statistics of the collective distribution. This is not a contradiction. It confirms the interpretation of $F_M^{(k)}$ as a resource diagnostic: when low-degree moments fail, one must supply additional readout resources, such as number resolution, parity sensitivity, tomography, randomized measurements, or interaction-based decoding.

\textit{Experimental readout bottlenecks and future tests.---}
The present results clarify a limitation that is common to many existing many-body metrology experiments.  Most platforms have become highly successful at preparing correlated probes, but the final measurement is often still organized around a small set of low-order collective observables: mean spin, spin noise, covariance matrices, number imbalance, or a few rotated collective components \cite{Leroux2010,Hosten2016,Bao2020,Bohnet2016,Linnemann2016,Mao2023,Bornet2023,Eckner2023}.  Such measurements are natural and powerful when the parameter response remains a smooth displacement or a Gaussian deformation of the collective state.  They are not, however, guaranteed to reveal the information carried by a strongly non-Gaussian many-body probe.  Our hierarchy gives a precise formulation of this limitation: a sensor may contain large QFI while the low-degree moment sector has small overlap with the SLD.  In that regime, an experiment can fail to observe the predicted gain not because the useful quantum state was not created, but because the readout projects onto the wrong part of operator space.

This distinction is especially important for nonlinear protocols, interaction-generated non-Gaussian states, Rydberg and optical-clock arrays, cavity-mediated ensembles, trapped ions, and spinor condensates, where the dynamics can generate fine angular structures on the collective Bloch sphere \cite{Pedrozo2020,Greve2022,Robinson2024,Marciniak2022,LewisSwan2024,Cao2024,Bluvstein2024,Schine2024}.  Conventional spin-squeezing or covariance measurements test only the first few layers of this structure.  A null or weak signal in these observables should therefore not be interpreted automatically as absence of metrological information.  It may instead indicate that the Fisher information has become hidden in high-rank collective correlations.  In this sense, the relevant experimental question is not only whether a many-body state has been prepared, but whether the apparatus has enough readout resolution to match the operator complexity of the optimal measurement.

The hierarchy suggests a direct experimental protocol.  One first estimates or bounds the QFI using existing certification, randomized measurement, or variational methods \cite{Cerezo2021,Elben2023,Choi2023,Oh2024,Haug2024,Vitale2024}.  One then reconstructs the accessible Fisher information from progressively higher collective moments, or from controlled families of rotated collective measurements.  If $F_M^{(1)}$ or $F_M^{(2)}$ nearly saturates $F_Q$, the sensor is readout-simple and standard collective detection is sufficient.  If instead $F_M^{(k)}\ll F_Q$ until $k$ becomes large, the experiment is readout-limited: the missing metrological gain is stored in correlations that ordinary collective readout cannot resolve.  The scale $k_{90}$ therefore acts as an experimentally meaningful ``readout depth'', analogous to a resolution requirement for the measurement rather than a property of state preparation alone.

This viewpoint also reinterprets advanced readout methods.  Interaction-based readout, phase magnification, and time reversal \cite{Strobel2014,Davis2016,Hosten2016b,Nolan2017,Macri2016,Colombo2022} should be understood as physical decoders that rotate a high-rank SLD back toward a low-degree observable before detection.  Parity-sensitive and number-resolved measurements \cite{Gessner2019,Kaubruegger2021} provide a different route: rather than simplifying the SLD, they measure finer features of the collective distribution and thereby access higher-order information directly.  In the language of the present work, a successful decoder or high-resolution measurement closes the gap between $F_Q$ and $F_M^{(1)}$ or $F_M^{(2)}$.  Thus, future experiments can test hidden Fisher information by measuring the same state before and after an echo, time-reversal, or interaction-based decoding sequence: closing the hierarchy gap would directly demonstrate the conversion of high-rank metrological information into a readable low-degree signal.

\textit{Conclusion.---}
We have shown that many-body quantum sensors can possess large QFI while rendering most of it inaccessible to standard low-order collective moment readout. This separation is not captured by the QFI alone, because the QFI optimizes over all possible measurements and therefore hides the cost of implementing the optimal readout.

To quantify this cost, we introduced a moment Fisher hierarchy $F_M^{(k)}$ and the associated readout-complexity scale $k_{90}$. In the driven collective spin studied here, first moments access only a percent-level fraction of the QFI by $N=100$, while recovering $90\%$ of it requires a moment degree $k_{90}\simeq28$. The rank-resolved SLD calculation independently shows that nonlinear dynamics transfers the locally optimal readout toward the highest available $SU(2)$ tensor ranks and suppresses its low-rank weight to the $10^{-3}$ scale under weak damping.

Our results show that metrological usefulness is determined not only by how much QFI a many-body state contains, but also by whether the available measurement resolves the operator sector in which that information resides.  Hidden Fisher information therefore turns the usual question of quantum metrology into a joint state-readout problem: a many-body probe may be highly sensitive, yet experimentally silent, until an appropriate decoder or high-order measurement is supplied.  The moment Fisher hierarchy provides a quantitative diagnostic of this accessibility and identifies readout complexity as a genuine bottleneck in many-body quantum sensing.

\textit{Data availability.---}
The numerical data and analysis scripts used to generate the figures are available from the authors upon reasonable request.

\bibliographystyle{apsrev4-2}
\bibliography{ref}

% ============================================================
% SUPPLEMENTAL MATERIAL
% ============================================================

\clearpage
\onecolumngrid

\section*{Supplemental Material: Hidden Fisher Information in Many-Body Quantum Sensors}

\section{Moment hierarchy and endpoint completeness}

This section gives the operator-space structure underlying the moment Fisher hierarchy used in the main text. The system is restricted to the permutation-symmetric spin-$j=N/2$ irreducible representation, whose Hilbert-space dimension is
\begin{equation}
d=N+1.
\end{equation}
The collective spin matrices are represented in the $J_z$ basis with magnetic quantum numbers
\begin{equation}
m=-j,-j+1,\ldots,j.
\end{equation}

The real moment space ${\cal M}_k$ consists of traceless Hermitian collective spin polynomials of degree at most $k$. Equivalently, in the spin-$j$ irreducible representation, the collective operator algebra decomposes into irreducible $SU(2)$ tensor ranks
\begin{equation}
\ell=0,1,\ldots,2j=N.
\label{eq:S_rank_range}
\end{equation}
The rank-$\ell$ sector has dimension $2\ell+1$. Therefore, the full traceless operator space has dimension
\begin{equation}
\sum_{\ell=1}^{N}(2\ell+1)
=
(N+1)^2-1
=
N(N+2).
\label{eq:S_full_dimension}
\end{equation}
Collective spin polynomials of degree $k$ generate tensor ranks up to $\ell=k$. Therefore, collective spin polynomials of degree $N$ generate the full traceless operator algebra,
\begin{equation}
{\cal M}_N
=
\{\mathrm{all\ traceless\ Hermitian\ operators\ on\ the\ spin}\text{-}j\ \mathrm{manifold}\}.
\label{eq:S_MN_complete}
\end{equation}
This establishes the finite-$N$ completeness of the hierarchy.

The endpoint completeness is essential for interpreting the results. The SLD $L_\alpha$ is Hermitian but need not be Hilbert--Schmidt traceless. This does not affect the hierarchy, because adding a multiple of the identity to an estimator changes neither its variance nor its parameter response, since $\mathrm{Tr}(\partial_\alpha\rho_\alpha)=0$. We may therefore use the traceless representative
\begin{equation}
\widetilde L_\alpha
=
L_\alpha
-
\frac{\mathrm{Tr}(L_\alpha)}{d}\,\mathbb{I},
\end{equation}
which belongs to ${\cal M}_N$. Measuring $L_\alpha$ locally saturates the QFI, and the identity shift is operationally irrelevant; hence the complete hierarchy satisfies
\begin{equation}
F_M^{(N)}=F_Q,
\qquad
\eta_N=\frac{F_M^{(N)}}{F_Q}=1.
\label{eq:S_endpoint}
\end{equation}
In the numerical calculations, this endpoint recovery is explicitly verified for every hierarchy curve used in the figures. This check distinguishes hidden Fisher information from numerical loss of QFI. If $\eta_N\simeq1$ but $\eta_k\ll1$ for small $k$, the information is present in the state but inaccessible to low-degree moment estimators.

\section{Optimization over moment estimators}

For a scalar moment estimator
\begin{equation}
X=\sum_\mu c_\mu O_\mu\in{\cal M}_k,
\label{eq:S_estimator}
\end{equation}
where $\{O_\mu\}$ is a Hilbert--Schmidt orthonormal basis of ${\cal M}_k$, the signal response vector is
\begin{equation}
g_\mu
=
\mathrm{Tr}
[
(\partial_\alpha\rho_\alpha)O_\mu
],
\label{eq:S_g}
\end{equation}
and the covariance matrix is
\begin{equation}
\Sigma_{\mu\nu}
=
\frac{1}{2}
\mathrm{Tr}
[
\rho_\alpha
\{\Delta O_\mu,\Delta O_\nu\}
],
\qquad
\Delta O_\mu=O_\mu-\langle O_\mu\rangle .
\label{eq:S_covariance}
\end{equation}
The error-propagation Fisher information associated with $X$ is
\begin{equation}
F_X
=
\frac{
(\partial_\alpha\langle X\rangle)^2
}{
(\Delta X)^2
}
=
\frac{
(c^Tg)^2
}{
c^T\Sigma c
}.
\label{eq:S_error_propagation}
\end{equation}
Maximizing over the real coefficient vector $c$ gives
\begin{equation}
F_M^{(k)}
=
g_k^T\Sigma_k^+g_k,
\qquad
\eta_k
=
\frac{F_M^{(k)}}{F_Q},
\label{eq:S_FMk}
\end{equation}
where $\Sigma_k^+$ is the Moore--Penrose pseudoinverse.

If the covariance matrix is nonsingular, Eq.~\eqref{eq:S_FMk} reduces to $g_k^T\Sigma_k^{-1}g_k$. If null directions are present, the pseudoinverse gives the correct variational result provided covariance-null directions are handled consistently. In the mixed-state calculations, the covariance matrix is diagonalized and only positive eigenvalue directions are retained. This avoids artificial enhancement from numerically singular directions.

The readout-complexity scale used throughout the paper is
\begin{equation}
k_{90}
=
\min\{k:\eta_k\geq 0.9\}.
\label{eq:S_k90}
\end{equation}
Small $k_{90}$ means that the QFI is directly visible to low-order collective moments. A large and system-size-dependent $k_{90}$ means that the QFI is present but distributed over increasingly high-degree collective correlations.

The hierarchy should be interpreted as an accessibility hierarchy for moment estimators, not as a restricted-POVM optimization. A full number-resolved histogram, a parity measurement, a randomized measurement protocol, or an interaction-based decoder may access information beyond $F_M^{(k)}$ at fixed low $k$. This distinction is important because the hierarchy is designed to diagnose when such additional readout resources become necessary.

\section{Efficient pure-state hierarchy for large $N$}

The large-$N$ calculations in the main text use unitary kicked-top dynamics and exploit a pure-state version of the hierarchy. This allows system sizes up to $N=100$ to be treated without constructing the full mixed-state covariance matrix in the $N(N+2)$-dimensional traceless operator space.

The initial state is the coherent state defined in Eq.~(2) of the main text. The state $|\psi_\alpha\rangle$ and its parameter derivative $|\partial_\alpha\psi_\alpha\rangle$ are propagated exactly through one eight-kick cycle. For a pure state, the QFI is
\begin{equation}
F_Q
=
4
\left(
\langle\partial_\alpha\psi|\partial_\alpha\psi\rangle
-
|\langle\psi|\partial_\alpha\psi\rangle|^2
\right).
\label{eq:S_pure_QFI}
\end{equation}
Define the tangent component orthogonal to the state,
\begin{equation}
|\partial_\alpha\psi_\perp\rangle
=
|\partial_\alpha\psi\rangle
-
|\psi\rangle
\langle\psi|\partial_\alpha\psi\rangle .
\label{eq:S_orthogonal_tangent}
\end{equation}
For a Hermitian observable $O$, define
\begin{equation}
|v_O\rangle
=
\Delta O|\psi\rangle,
\qquad
\Delta O=O-\langle O\rangle .
\label{eq:S_vO}
\end{equation}
The accessible Fisher fraction $\eta_k$ is then the squared length of the projection of $|\partial_\alpha\psi_\perp\rangle$ onto the real span of the vectors $|v_O\rangle$ generated by all $O\in{\cal M}_k$, divided by $\|\partial_\alpha\psi_\perp\|^2$.

This pure-state construction is numerically stable because the tangent space can be built rank by rank using an irreducible $SU(2)$ tensor basis. Working in the tensor basis avoids the severe ill conditioning that arises from high-degree monomials in $J_x$, $J_y$, and $J_z$. The endpoint check $\eta_N\simeq 1$ is maintained up to $N=100$, confirming that the hierarchy recovers the QFI when the full operator space is included.

\section{Detuning scan of the strong-shear readout barrier}

The main text focuses on the resonant nonlinear kick $\beta=\pi j$. To test whether the readout barrier is an artifact of exact resonance, we scan the scaled nonlinear strength
\begin{equation}
r=\frac{\beta}{\pi j}.
\label{eq:S_scaled_twist}
\end{equation}
All other parameters are kept fixed: the initial coherent state has $\Theta=\pi/3$ and $\Phi=0$, the estimated rotation is evaluated at $\alpha=\pi/2$, and the evolution consists of one eight-kick unitary cycle.

\begin{figure*}[t]
\centering
\safeincludegraphics[width=0.98\textwidth]{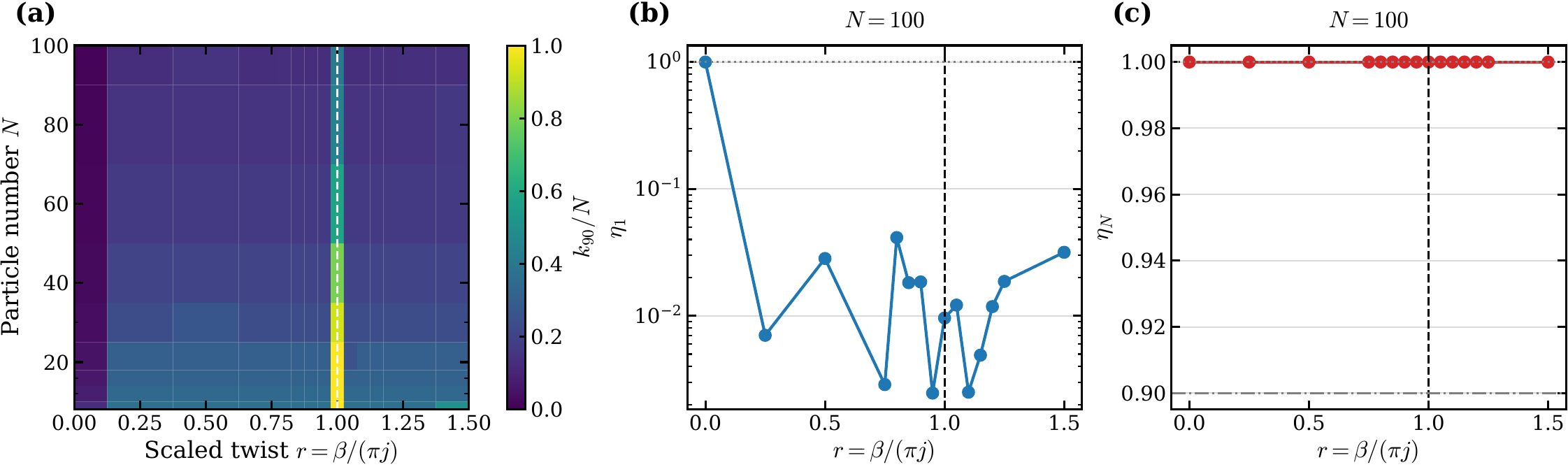}
\caption{\textbf{Detuning scan of the nonlinear readout barrier.}
The scaled nonlinear strength $r=\beta/(\pi j)$ is scanned around the resonant strong-shear point.  The scan shows that low-degree accessibility is suppressed throughout a broad nonlinear regime while the full-degree endpoint remains recovered.}
\label{fig:S_detuning_scan}
\end{figure*}

For each value of $r$ and $N$, we compute the QFI, the hierarchy $\eta_k=F_M^{(k)}/F_Q$, and the readout degree $k_{90}$. Figure~\ref{fig:S_detuning_scan} shows the resulting detuning scan. The linear case $r=0$ has no appreciable readout barrier: the QFI is already accessible from the lowest collective moments. Once nonlinear kicking is introduced, the linear accessibility $\eta_1$ is strongly suppressed. At $N=100$, $\eta_1$ lies at the percent level or below for much of the nonlinear regime. Thus, ordinary collective-spin readout fails well beyond the exactly resonant point.

The strongest readout-complexity barrier occurs near the resonant strong-shear point $r\simeq 1$, where the normalized readout degree is strongly enhanced relative to the linear limit. The endpoint recovery remains $\eta_N\simeq 1$ throughout the scan. This verifies that the QFI is not destroyed by the nonlinear dynamics and is not lost numerically. Instead, it is redistributed into high-degree collective moment sectors.

The shear mechanism gives the relevant physical scale. The classical twist angle is
\begin{equation}
\chi(s)=\beta s_y,
\label{eq:S_twist_angle}
\end{equation}
so that
\begin{equation}
\frac{\partial\chi}{\partial s_y}
=
\beta
=
r\pi j.
\label{eq:S_twist_gradient}
\end{equation}
Near $r=1$, the shear strength is $O(N)$. This efficiently transfers the parameter tangent and the SLD into high-rank tensor sectors. Since degree-$k$ collective moments resolve ranks only up to $\ell=k$, recovering the QFI then requires high-degree moment information.

The detuning scan supports two conclusions. First, the suppression of low-order accessibility is a broad feature of nonlinear dynamics, not an isolated numerical accident. Second, the readout barrier is strongest near the resonant strong-shear region, consistent with the fact that the local twist gradient $\partial\chi/\partial s_y=\beta$ grows with $N$ at fixed $r\simeq1$.

\section{Finite-size detuning curves}

\begin{figure}[t]
\centering
\safeincludegraphics[width=0.4\linewidth]{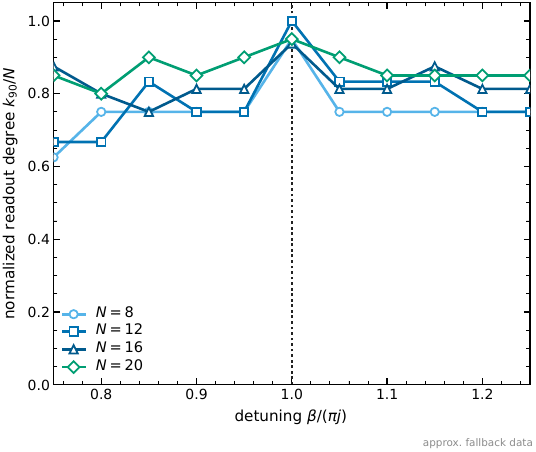}
\caption{\textbf{Finite-size detuning curves.}
Normalized readout complexity $k_{90}/N$ as a function of detuning for the smaller system sizes used to resolve the strong-shear window.  The vertical guide marks $\beta/(\pi j)=1$.}
\label{fig:S_finite_detuning}
\end{figure}

Figure~\ref{fig:S_finite_detuning} shows the normalized readout complexity $k_{90}/N$ as a function of detuning for several smaller system sizes. This representation complements the large-$N$ heat-map scan in Fig.~\ref{fig:S_detuning_scan} by resolving individual finite-size curves near the resonant strong-shear point.

The curves show that the readout barrier is not a single isolated outlier. For all displayed sizes, $k_{90}/N$ remains strongly enhanced across the strong-shear detuning window and reaches its largest values near $\beta/(\pi j)=1$. The precise detuning dependence is not perfectly monotonic at small $N$, as expected for a finite-dimensional Floquet system. The robust feature is the persistence of a large finite-size readout degree across a finite detuning interval, rather than the detailed location of each fluctuation.

This analysis is useful because it separates two issues that can otherwise be confused. The exact value of $k_{90}$ at a particular detuning is a finite-size quantity. By contrast, the persistence of a large normalized readout degree across a finite detuning window reflects the underlying shear mechanism and supports the interpretation that high-degree collective correlations are required to recover the QFI.

\section{Weak-damping mixed-state validation}

The large-$N$ calculations in the main text use unitary pure-state evolution in order to reach $N=100$. To verify that hidden Fisher information is not an artifact of ideal pure-state dynamics, we also compute the full mixed-state moment hierarchy with weak collective damping. A single noisy Floquet step is
\begin{equation}
\rho
\mapsto
U_{\alpha,\beta}
\left[
e^{{\cal L}_d}\rho
\right]
U_{\alpha,\beta}^{\dagger},
\label{eq:S_noisy_step}
\end{equation}
where
\begin{equation}
{\cal L}_d(\rho)
=
\gamma
\left(
2J_-\rho J_+
-
\{J_+J_-,\rho\}
\right),
\qquad
\gamma=10^{-3}.
\label{eq:S_Lindblad}
\end{equation}
For the hierarchy validation summarized in Table~\ref{tab:S_weak_damping}, this noisy map is iterated for eight kicks, matching the depth of the large-$N$ unitary protocol. The mixed-state covariance matrix is diagonalized and inverted only on its positive-eigenvalue support.

The resonant protocol retains a large readout degree over the displayed sizes, while the $\beta=0$ baseline remains accessible at the lowest moment degree. In the resonant data, $\eta_1$ is strongly suppressed whereas the full-degree endpoint satisfies $\eta_N\simeq1$. Thus, weak damping does not remove the separation between stored QFI and low-order accessibility on the simulated sizes.

The mixed-state calculation is substantially more expensive than the pure-state projection method because it constructs the covariance matrix in the full $N(N+2)$-dimensional traceless operator algebra. We therefore restrict this validation to smaller systems. Its purpose is not to infer a large-$N$ scaling law at fixed dissipation, but to establish that the hierarchy gap survives beyond ideal unitary evolution.

Dissipation also has a clear physical role: it preferentially suppresses fine, high-rank structure and therefore acts as a low-pass filter on the collective Bloch sphere. At sufficiently large size and fixed nonzero damping, this effect can reduce the high-rank information itself. The present finite-size calculation only establishes that the readout barrier appears before weak damping eliminates the metrological response.

\begin{table}[t]
\centering
\caption{
Weak-damping finite-size hierarchy data for the resonant protocol, $\beta=\pi j$ and $\gamma=10^{-3}$, after eight successive noisy Floquet steps. The endpoint $\eta_N\simeq1$ confirms recovery of the QFI in the complete moment space.
}
\begin{tabular}{c c c c c}
\hline
$N$ & $k_{90}$ & $k_{90}/N$ & $\eta_1$ & $\eta_N$ \\
\hline
8  & 8  & 1.0000 & 0.0511 & 1.0000 \\
12 & 12 & 1.0000 & 0.0205 & 1.0000 \\
16 & 15 & 0.9375 & 0.0038 & 1.0000 \\
20 & 19 & 0.9500 & 0.0018 & 1.0000 \\
\hline
\end{tabular}
\label{tab:S_weak_damping}
\end{table}

\section{Adjoint-rank interpretation and tangent-space support}

The adjoint Casimir
\begin{equation}
{\cal C}_{\mathrm{ad}}(O)
=
\sum_{a=x,y,z}
[J_a,[J_a,O]]
\label{eq:S_adjoint_Casimir}
\end{equation}
provides a basis-independent notion of operator smoothness on the collective Bloch sphere. Its eigenoperators are the irreducible tensor operators $T_{\ell m}$,
\begin{equation}
{\cal C}_{\mathrm{ad}}(T_{\ell m})
=
\ell(\ell+1)T_{\ell m}.
\label{eq:S_tensor_eigenvalue}
\end{equation}
Low tensor ranks describe smooth collective structure, whereas higher ranks resolve progressively finer angular structure.

For an operator $O$, the Hilbert--Schmidt rank weight is
\begin{equation}
w_\ell(O)
=
\frac{
\sum_{m=-\ell}^{\ell}
|\mathrm{Tr}(T_{\ell m}^{\dagger}O)|^2
}{
\mathrm{Tr}(O^\dagger O)
}.
\label{eq:S_rank_weight}
\end{equation}
The main-text rank figure applies this decomposition to the SLD, because the SLD specifies the locally optimal measurement direction. For comparison with the moment hierarchy, the SLD is understood modulo an additive identity component, which is irrelevant to both response and variance. The nonlinear protocol shifts the readout-relevant SLD structure toward high tensor rank and strongly suppresses its low-rank weight.

\begin{figure*}[t]
\centering
\safeincludegraphics[width=0.92\textwidth]{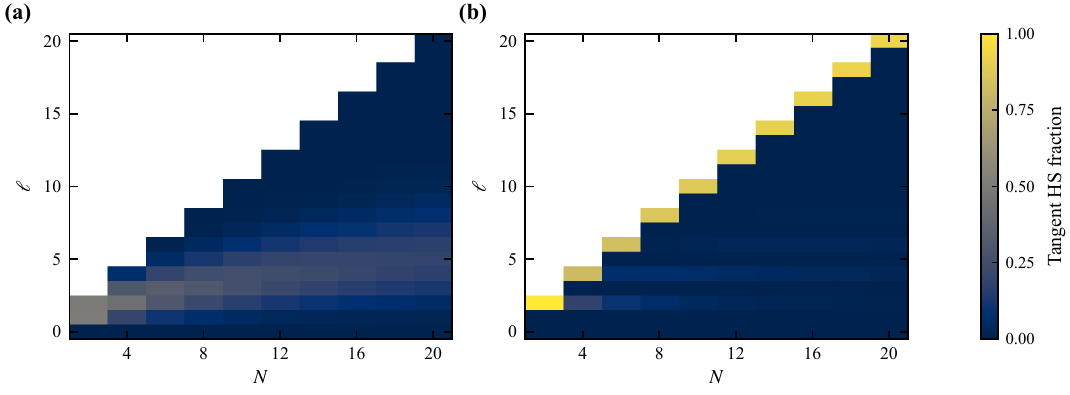}
\caption{\textbf{Tensor-rank structure of the parameter tangent.}
Hilbert--Schmidt rank weights $w_\ell(\partial_\alpha\rho_\alpha)$ for one weakly damped Floquet step with $\gamma=10^{-3}$.
(a) Linear protocol, $\beta=0$.
(b) Resonant nonlinear protocol, $\beta=\pi j$.
The nonlinear kick transfers the tangent toward substantially higher $SU(2)$ tensor ranks, approaching the upper rank boundary as $N$ increases. This complements the SLD-rank diagnostics in the main text and directly supports the tangent-space shear mechanism.}
\label{fig:S_tangent_rank}
\end{figure*}

To test the proposed \emph{mechanism} rather than only its optimal-readout consequence, we additionally decompose the parameter tangent $\partial_\alpha\rho_\alpha$. Figure~\ref{fig:S_tangent_rank} uses the same weak-damping parameters as the main-text rank analysis, $\gamma=10^{-3}$ and one noisy Floquet step, for $N=2,4,\ldots,20$. In the linear protocol the tangent occupies substantially lower normalized ranks, whereas the resonant nonlinear kick drives most of its Hilbert--Schmidt weight toward the upper part of the available rank spectrum. At $N=20$, for example, the mean normalized tangent rank is approximately $0.27$ for $\beta=0$ and $0.95$ for $\beta=\pi j$; the corresponding rank-$\ell\leq2$ weights are approximately $7.1\times10^{-2}$ and $7.8\times10^{-3}$.

This comparison is important because the tangent and the SLD play different roles. The tangent describes how the state changes with the parameter, while the SLD describes the locally optimal observable for converting that change into Fisher information. Their simultaneous migration toward high tensor rank in the nonlinear protocol supports the tangent-space shear interpretation without identifying the two operators.

\section{Relation to classical Fisher information of explicit collective measurements}

The moment hierarchy is not a restricted-POVM optimization. It optimizes the best scalar error-propagation estimator formed from collective moments up to degree $k$. By contrast, for a specified POVM $\{\Pi_x\}$, the classical Fisher information is
\begin{equation}
F_C[\{\Pi_x\}]
=
\sum_x
\frac{
[\partial_\alpha \mathrm{Tr}(\rho_\alpha\Pi_x)]^2
}{
\mathrm{Tr}(\rho_\alpha\Pi_x)
}.
\label{eq:S_classical_FI}
\end{equation}
The QFI is the supremum of $F_C$ over all POVMs. Therefore, $F_M^{(k)}$ should be interpreted as the Fisher information accessible to a specified low-degree moment estimator, not as the best possible Fisher information under every measurement compatible with collective hardware.

To make this distinction explicit, we compute the classical Fisher information for two concrete collective readouts. First, we consider a global rotation followed by number-resolved measurement in the $J_z$ basis. For a rotation $R(\theta,\phi)$, the probabilities are
\begin{equation}
p_m(\alpha;\theta,\phi)
=
|\langle m|R(\theta,\phi)|\psi_\alpha\rangle|^2,
\label{eq:S_rotated_probabilities}
\end{equation}
and the corresponding classical Fisher information is
\begin{equation}
F_C^{J_z}(\theta,\phi)
=
\sum_m
\frac{
[\partial_\alpha p_m(\alpha;\theta,\phi)]^2
}{
p_m(\alpha;\theta,\phi)
}.
\label{eq:S_Jz_FI}
\end{equation}
We optimize $F_C^{J_z}$ over a grid of global rotations.

Second, we coarse-grain the same rotated $J_z$ distribution into parity outcomes,
\begin{equation}
p_\pm
=
\sum_m
p_m
\frac{
1\pm(-1)^{j+m}
}{2},
\label{eq:S_parity_probabilities}
\end{equation}
and compute the optimized two-outcome parity Fisher information.

\begin{figure}[t]
\centering
\safeincludegraphics[width=0.4\linewidth]{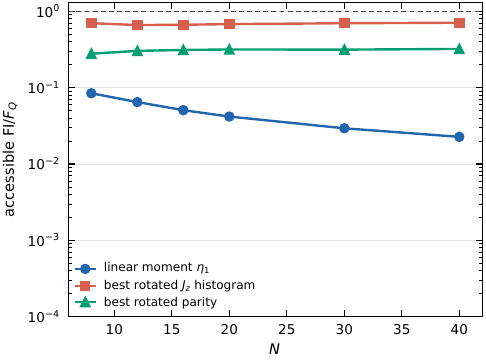}
\caption{\textbf{Classical Fisher information from explicit collective readouts.}
Optimized number-resolved collective histograms and parity-sensitive measurements can access more Fisher information than low-degree scalar moments, illustrating that the hierarchy diagnoses the resources required by a readout rather than the absence of QFI.}
\label{fig:S_collective_cfi}
\end{figure}

Figure~\ref{fig:S_collective_cfi} shows that number-resolved collective histograms and parity-sensitive measurements can recover substantially more information than linear moments. This is not a contradiction. Such readouts implicitly contain high-order statistics of the collective distribution and therefore exceed the low-degree scalar moment hierarchy at fixed $k$.

This comparison clarifies the operational meaning of hidden Fisher information. The suppression of $\eta_1$ diagnoses a limitation of low-order moment readout, not an absence of QFI and not a fundamental failure of all collective measurements. A full histogram, a parity measurement, or an active decoder supplies additional readout resources. The moment hierarchy quantifies when such resources become necessary.

\section{Semiclassical shear estimate}

The nonlinear kick induces the classical map
\begin{align}
s_y^+ &= s_y,\\
s_x^+ &= s_x\cos(\beta s_y)-s_z\sin(\beta s_y),\\
s_z^+ &= s_z\cos(\beta s_y)+s_x\sin(\beta s_y),
\end{align}
on the normalized collective spin $s=J/j$. The local twist angle is
\begin{equation}
\chi(s)=\beta s_y,
\label{eq:S_chi}
\end{equation}
and at the resonant kick $\beta=\pi j$,
\begin{equation}
\frac{\partial\chi}{\partial s_y}
=
\beta
=
\pi j
=
O(N).
\label{eq:S_resonant_shear}
\end{equation}
Thus, the local shear itself becomes stronger with system size at fixed resonant scaling. A smooth parameter displacement can consequently develop increasingly fine angular structure under a fixed-depth nonlinear sequence.

Tensor rank $\ell$ resolves angular structure on a scale of order
\begin{equation}
\delta\theta\sim \frac{1}{\ell}.
\end{equation}
If the shear creates features whose angular scale decreases with $N$, increasingly high tensor ranks are required to resolve them. Because degree-$k$ collective polynomials contain tensor ranks only up to $\ell=k$, the same mechanism naturally produces a growing moment-readout depth.

This is a physical scaling estimate for the generated operator structure, not a proof that $k_{90}/N$ approaches a nonzero constant. The numerical hierarchy determines $k_{90}$ directly. Over the simulated range it shows substantial growth of the absolute readout degree, while the rank-resolved SLD and tangent diagnostics independently confirm migration toward high-rank sectors.

\section{Finite-size trend of $k_{90}/N$}

The large-$N$ unitary data in the main text show that the absolute degree required to recover $90\%$ of the QFI grows substantially over the accessible system sizes; at $N=100$ the resonant calculation gives $k_{90}\simeq28$. Because the normalized quantity $k_{90}/N$ displays visible finite-size crossover, we do not infer a unique asymptotic law from the available sizes. Instead, we compare the data with two simple diagnostic forms,
\begin{equation}
\frac{k_{90}}{N}
=
c+aN^{-b},
\label{eq:S_fit_nonzero}
\end{equation}
and
\begin{equation}
\frac{k_{90}}{N}
=
aN^{-b}.
\label{eq:S_fit_zero}
\end{equation}
Equation~\eqref{eq:S_fit_nonzero} permits a nonzero limiting fraction $c$, whereas Eq.~\eqref{eq:S_fit_zero} permits the normalized degree to vanish.

\begin{figure*}[t]
\centering
\safeincludegraphics[width=0.98\textwidth]{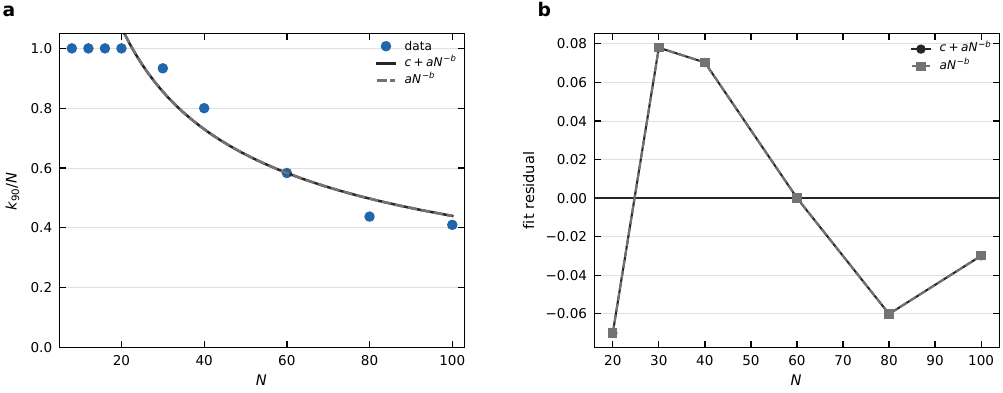}
\caption{\textbf{Finite-size trend of normalized readout complexity.}
The resonant data are compared with nonzero- and zero-asymptote fitting forms. The fits are used only to illustrate finite-size crossover and are not interpreted as establishing a unique large-$N$ limit.}
\label{fig:S_scaling_extrapolation}
\end{figure*}

Figure~\ref{fig:S_scaling_extrapolation} compares these forms with the finite-size data and their residuals. The purpose is diagnostic: the accessible sizes are insufficient to discriminate a nonzero limiting fraction from a slowly decreasing normalized degree with asymptotic confidence. The robust conclusion used in the main text is therefore finite-size and operational: low-order moments remain strongly insufficient, the absolute readout degree grows to high order, and the full hierarchy eventually recovers the QFI.

\section{Experimental observables and reconstruction}

Degree one consists of the collective spin components $J_x$, $J_y$, and $J_z$. Degree two includes second moments, spin-squeezing observables, and spin-noise covariances. Higher degrees correspond to higher-order cumulants or symmetrized collective-spin products. Depending on the platform, these quantities can be accessed through repeated collective measurements after global rotations, spin-noise spectroscopy, randomized measurements, parity-sensitive detection, number-resolved histograms, or interaction-based decoding.

The hierarchy separates limitations of state preparation from limitations of readout. If an independent calculation or certification indicates large QFI while $F_M^{(1)}$ and $F_M^{(2)}$ remain much smaller, then the missing sensitivity lies outside the low-degree moment sector. A direct experimental test is therefore to compare the same probe before and after a richer readout operation. Recovery of Fisher information after an echo, time-reversal sequence, interaction-based decoder, parity-sensitive measurement, or number-resolved measurement would demonstrate conversion of high-complexity metrological information into a readable signal.

The hierarchy does not rank these readout strategies by hardware cost. Its role is to provide a common operational benchmark: a readout architecture is successful when it closes the gap between the Fisher information available in the state and that accessible to the initially restricted low-degree observables.

\section{Numerical robustness and safeguards}

\begin{figure*}[t]
\centering
\safeincludegraphics[width=0.98\textwidth]{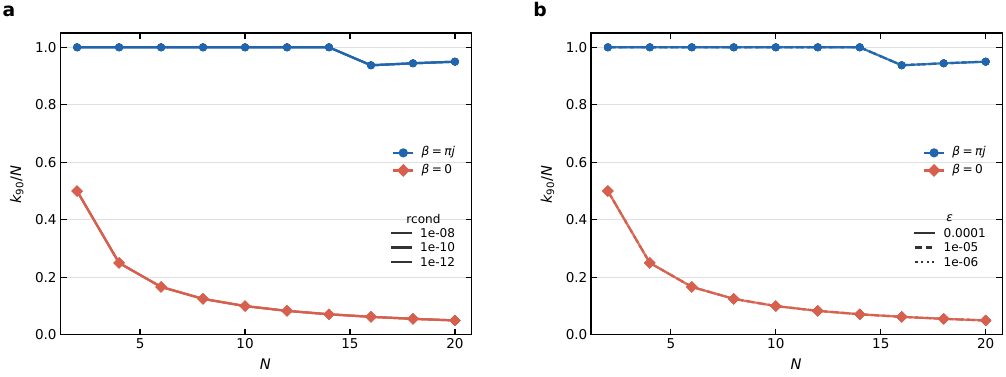}
\caption{\textbf{Numerical robustness checks.}
The resonant readout complexity is stable under changes of the covariance pseudoinverse cutoff and finite-difference step over the ranges used in the calculations.}
\label{fig:S_numerical_robustness}
\end{figure*}

Figure~\ref{fig:S_numerical_robustness} shows stability against numerical choices. The covariance pseudoinverse cutoff is varied between $10^{-8}$ and $10^{-12}$, and finite-difference checks use steps between $10^{-4}$ and $10^{-6}$. These checks cover the derivative evaluation used in the rank-resolved analysis as well as the hierarchy diagnostics; the conclusions are unchanged within the numerical resolution relevant here.
\noindent

The following safeguards are enforced in all hierarchy calculations:

\begin{enumerate}
\item At $k=N$, the moment basis contains $N(N+2)$ traceless operators, equivalently all nonzero $SU(2)$ tensor ranks $\ell\leq N$.

\item The endpoint recovery $\eta_N\simeq 1$ is verified for every hierarchy curve used in the figures.

\item In the mixed-state calculation, covariance inversion uses a positive-semidefinite eigenvalue pseudoinverse.

\item Accessibility curves use unclipped $\eta_k$ values for diagnostics.

\item The equality $F_M^{(N)}\simeq F_Q$ is checked independently through the plotted endpoint Fisher information.

\item Tangent propagation is checked against finite differences on representative system sizes and parameters.
\end{enumerate}

These checks ensure that the observed suppression of low-degree accessibility is not caused by basis incompleteness, covariance singularities, finite-difference errors, or numerical loss of QFI. The hierarchy endpoint always recovers the QFI, confirming that the information is hidden from low-degree moment readout rather than absent from the state.

\end{document}